\documentclass[12pt]{article}
\input epsf.sty
\newcommand{\beq}{\begin{equation}}
\newcommand{\eeq}{\end{equation}}
\newcommand{\bra}{\begin{array}}
\newcommand{\era}{\end{array}}
\newcommand{\be}{\beta}
\newcommand{\al}{\alpha}
\newcommand{\ga}{\gamma}

\newcommand{\Om}{\Omega}

\newcommand{\si}{\sigma}

\author{J. Douari \footnote{douari@sun.ac.za} \\
\small\it Stellenbosch Institute for Advanced Study, Private Bag X1,\rm\\
\small\it Matieland, Stellenbosch, 7601, South Africa\rm }
\title{$\phantom{~~}$Exotic Particles and $w_\infty$-Algebras in Two- and High-Dimensional Spaces}
\begin{document}
\maketitle
\vspace{3cm}
\section*{Abstract}
\hspace{.3in}We construct a set of noncommuting translation operators in two and high-dimensional lattices. The algebras they close are $w_{\infty}$-algebras. The construction is based on the introduction of noncommuting elementary link operators which link two neighborhood sites in the lattice. This kind of operators preserve the braiding nature of exotic particles living basically in two-dimensional space. 

\newpage
\section{Introduction}
\hspace{.3in}Exotic particles are known as excitations and quasi-particles or anyons; i.e. fermions (bosons) carrying
odd (even) number of elementary magnetic flux quanta \cite{quanta}. They are living in two-dimensional space as composite particles having arbitrary spin, and they are characterized by fractional statistics which is interpolating between bosonic statistics and fermionic one \cite{anyon}. Several works were done to find out their quantum theory and its right model is not yet reached. One of the field-theories describing anyons is the model, where the matter is interacting with the Chern-Simons (CS) gauge field \cite{GCS}. In the reference \cite{GConn}, Stern has introduced another approach to treat anyons that does not require the CS term, but introduces a generalized connection with which the conserved U(1) current is coupled in a gauge invariant way \cite{Any}. In this model the gauge field is dynamic and the potential has the confining nature which makes the model different \cite{GC}. Another interesting investigation concerning anyons is finding their quantum symmetries \cite{Sym}. This constitutes the propose of this letter. We show that $w_{\infty}$-algebras characterize the noncommutative or braiding nature of exotic particles.\\ 

In this letter we shall focus on a special case of w-algebras containing an infinite link operators constructing general translation generators in two and high-dimensional lattices. We choose the spacing equals to one. As known in the literature, the $W_N$ algebras have currents of spins 2, 3, . . . ,N, for each of which there is a non-trivial central extension \cite{Walg}. These algebras are non-linear, but there is an $N \longrightarrow\infty$ limit, known as $w_\infty$, in which linearity is regained \cite{Winfty}. The algebra $w_\infty$ is a classical limit $N \longrightarrow\infty$ of $W_N$; we expect that it contains generators of all spins $s > 2$. In our case the spins are arbitrary; $s\in\bf R\rm$. The $w_\infty$ algebra is also a particular generalization of the Virasoro algebra with generators of higher spin 2, 3, ...\\

The main goal of this work is to understand the way the exotic particles could move in high-dimensional space and not only two-dimensional one. We first suggest the discretization of two-dimensional space. We use the link operators as elementary translation operators which will play a main role in this letter. We assume the noncommutativity of these operators which reflects the behaviour of anyons in two-dimensional space; i.e. the noncommuting link operators lead to Yang-Baxter equation (YBE) from which we get the braiding equation for anyons. This matter constitutes the subject of the second section. In the third section, the link operators are the background again once to define the translation operators in two-dimensional lattice generating the quasi-particles motions. The obtained generators close a $w_\infty$ algebra characterizing anyons. Actually, our real space is not two dimensions, so what about the motion of exotic particles in high dimensions? Thus, we suggest the discretization of high-dimensional space and the result is a lattice equals to tensor product of infinite or $d$ times ($d>2$) two-dimensional lattices. The particles move in this high-dimensional lattice from one two-dimensional lattice to another by using the links connecting two neighborhood sites. This is done mathematically by applying link operators. The suggestion of their noncommutative case leads us to define a general translation operators which close a $w_\infty$ algebra characterizing exotic particles in high-dimensional lattice and their braiding properties are conserved. 
\section{Link Operators on two-dimensional lattice}
\hspace*{.3in}This section is devoted to introduce noncommuting operators denoted by $L_\ell$ which allow the transition
from one site to an arbitrary other one on a given two-dimensional lattice. We begin with the definition of the above operators and we show that the associativity of the algebra generated by the non-commuting elements $L_\ell$ is equivalent to
Yang-Baxter equation.\\

Let's denote the two-dimensional lattice by $\Omega$ and we define the link
operators $L_\ell$ as follows:
\beq
\phi^\ell_i(x)\equiv L_\ell\phi_i(x)\eeq
where $\phi_i(x)/i=1,\dots,d$ is a $d$-dimensional vector function on $\Omega$,
$x$ is defined to be a couple of two integer $x\equiv (x_1,x_2)$ ($x_1$ and
$x_2$ are respectively the horizontal and longitudinal coordinates of a given
site $x$ of $\Omega$).
\vspace{1cm}
$$\begin{picture}(0,0)
\put(-30,0){\line(0,-1){90}}
\put(20,0){\line(0,-1){90}}
\put(-50,-20){\line(1,0){90}}
\put(-50,-70){\line(1,0){90}}
\put(-33,-23){$\bullet$}
\put(-33,-73){$\bullet$}
\put(17,-23){$\bullet$}
\put(17,-73){$\bullet$}
\put(-10,-23){$\leftarrow$}
\put(-10,-73){$\rightarrow$}
\put(17,-50){$\uparrow$}
\put(-33,-50){$\downarrow$}
\put(-15,-15){$L_{-1}$}
\put(-15,-80){$L_{1}$}
\put(-52,-45){$L_{-2}$}
\put(25,-45){$L_{2}$}
\put(-68,-80){$(x_1,x_2)$}
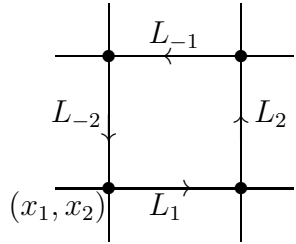
\put(-200,-140){{\bf Fig.1} -- Elementary plaquette $P_x$, $x=(x_1,x_2)\in\Omega$, on which}
\put(-200,-160){$\phantom{~~~~~~~~~~}$ the particle is translated from $x$ to $x$.}
\end{picture}$$
\vspace{6cm}

In the equation (1) $\ell=\pm 1,\pm 2$ are the four possible orientations
on $\Omega$ (Fig.1), so the index ``$\ell$'' indicates a moving along the
direction $\ell$ on $\Omega$. Thus the $\phi^\ell_i(x)$ are defined as follows
\beq\bra{ll}
\phi^1_i(x)\equiv \phi_i(x_1+1, x_2),\quad \phi^2_i(x)\equiv
\phi_i(x_1,x_2+1)\\\\
\phi^{-1}_i(x) \equiv \phi_i(x_1-1,x_2),\quad \phi^{-2}_i(x)\equiv
\phi_i(x_1,x_2-1)\era\eeq

We note that in these definitions, the element $L_\ell$ is regarded as an
operator linking two neighbouring lattice sites, so we call it the link operator. In the
simple case where these operators commute, one can write
\beq
L^{(\ell')}_\ell\cdot L_{\ell'}=L^{(\ell)}_{\ell'}\cdot L_\ell\eeq
we point out that in this relation, we adopt the notation $L^{(\ell')}_\ell$
for which we suppose that the link operator $L^{(\ell')}_\ell$ acts on a given
vector function $\phi_i (x)$ in the direction $\ell$ by keeping invariant the direction described
by $\ell'$.\\

The product ``$\cdot$'' in Eq.(3) is simply the composition of the operators
$L_\ell$'s, this composition occurs from the definition (1). Now, by
introducing a
matrix denoted by $R_{\ell\ell'}$, we break the commutativity of the product (Eq.(3)) as
follows
$$R_{\ell\ell'}:V\otimes V\to V\otimes V$$
and
\beq
(R_{\ell\ell'})^{ij}_{mn} \left( L^{(\ell')}_\ell\right)^m_{m'}\cdot
\left(L_{\ell'}\right)^n_{n'}=
\left(L^{(\ell)}_{\ell'}\right)^j_{n'}\cdot (L_\ell)^i_{m'}\eeq
where $V$ is a $d$-dimensional vector space in which lives the multi-component
functions ``$\phi_i$''. The indices $i$, $j$, $m'$ and $n'$ take the values
$1,2,\dots,d$.\\

The equation (4) can be rewritten in a compact form as
\beq(R_{\ell\ell'})_{12}\left((L^{(\ell')}_\ell)_1\cdot
(L_{\ell'})_2\right)=(L^{(\ell)}_{\ell'})_2\cdot (L_\ell)_1\eeq
owing to  this equality, one can prove, by a direct computation that the
relation
\beq(R_{\ell\ell'})_{12}(R_{\ell'\ell})_{21}=1\!1\ \otimes 1\!1\ \eeq
where $1\!1$ is the unit $d\times d$ matrix acting on $V$.\\

Starting from the above tools and by requiring the product in Eq.(3) to be associative, we obtain the well known
Yang-Baxter equation (YBE) on $(R_{\ell\ell'})$
\beq
(R_{\ell\ell'})_{12}(R^{(\ell')}_{\ell\ell''})_{13}(R_{\ell'\ell''})_{23}=(R^{
(\ell)}_{\ell'\ell''})_{23}
(R_{\ell\ell''})_{13}(R^{(\ell'')}_{\ell\ell'})_{12}\eeq

We recall that in the literature this equation is seen as
a representation of the braid group. The latter plays for the intermediate
quantum statistics the same role played by the permutation group for bosonic
and fermionic statistics. In mathematical sense, it suffices to multiply the
matrix $R$ in (7) by a permutation one $P$ as
$$B=P\cdot R$$
and the YBE becomes:
\beq
B_{12} B_{23}B_{12}=B_{23}B_{12}B_{23}\eeq
We notice that in the classical limit $R=1\!1 \ \otimes 1\!1$,
this equation becomes trivial. The equality (8), known as the braid relation appears in the study of
intermediate statistics (especially the anyonic ones). Consequently, The noncommuting link operators seem to be an interesting objects to treat the two-dimensional system and specially to lead with their symmetries. This matter constitutes the purpose of the next section.
\section{$ w_\infty $-algebras in two and high-dimensional lattices}
\hspace{.3in}By using the above mathematical tools, we construct an algebra in d-dimensional lattice describing the translations of particles through lattice's links applying elementary link operators. As above we consider the discretization of 2-dimensional space and we use the so-called link operators $L_\ell $ as elementary translation operators, such that
\beq
L_\ell |x\rangle =|x+\ell \rangle,
\eeq
with $|x\rangle $ is a state of a system living on 2-dimensional lattice at the site $"x"$ (fig.1).
By using the $L_\ell $ we can translate a state $|x\rangle $ from the site $"x"$ to the same site on $\Om$ along of an elementary plaquette $P_x$ for example.\\

We consider the non-commutative elementary translation operators defined by (4) and we choose the following notation
\beq
L_i L_j =R_{ij} L_j L_i ,\phantom{~~~}i,j=\pm 1,\pm 2,
\eeq
with the following propose
\beq
R_{ij} =e^{i\epsilon_{ij}\al}, \phantom{~~~~}\epsilon_{ij}=\pm 1
\eeq
where $\al$ is a real parameter depending on the statistics of the considered system and $\epsilon_{ij}$ depending on the direction of rotation on the lattice; + for the anticklocwize direction and - for its opposite.\\

To construct a quantum symmetry as Fairke-Fletcher-Zachos (FFZ) algebra in two-dimensional lattice, we define a noncommuting translation operators $T_x$ as
\beq
T^n_\ell=R_{ij}^{n\ell \over 2} L^{n}_{\pm i}L^{\ell}_{\pm j}\eeq
where $n,\ell\in\bf{N}\rm$, $i,j=1,2$ and $i\not= j$.\\

The operator $L_\ell$ allows the transition from the site $x_i$ to the site $x_{i+\ell}$ on $\Omega$, $\ell=\pm1,\pm2$ as seen in Fig.1, and the generators $T^n_\ell$ are regarded as translations made by $n$ times of elementary translation operators $L_{\pm i}$ and $\ell$ times of $L_{\pm j}$ in $\Omega$ with $i,j=1,2$. We require that the product of these operators is given by 
\beq
T^n_\ell T^m_k =K(n,\ell;m,k)T^{n+m}_{\ell+k}\eeq
The motivation of this choice of the product between these two translation
operators is due to the fact that the link operators do not commute and thus the composition of
two translations must lead naturally to a translation.\\

Before giving the complete description of the FFZ algebra, we notice that in the literature this
algebra has been poorly realized in a mathematical way. So, one of the main results of this work
is to construct the FFZ algebra on the lattice $\Omega$ and its generalization in high-dimensional lattice.
We suggest that the translation operators $T^n_\ell$ do not commute.This non-commutativity property is described by the introducing of some matrix $R$ as
$$T^n_\ell T^m_k=R^{ab}_{\ell k} T^m_aT^n_b .$$
Following the above assumption and requiring that the translation operators don't commute, we prove by a direct calculation that the parameters $K(n,\ell;m,k)$ appearing in (13) have the expression
$$ K(n,\ell;m,k)=e^{i\alpha_{2}(n,\ell;m,k)\over  (m,k)\times (n,\ell)}.$$
The function $\alpha_2 (n,\ell;m,k)$ depending on the integer numbers $n,\ell,m$ and $k$ is introduced as
$$\alpha_2:\bf{N}\rm^2\times\bf{N}\rm^2\to\bf{N}\rm.$$ 

Consequently, we find that the elements $T$'s are nothing but the generators of FFZ algebra on the lattice $\Omega$, we have then
\beq
[T^n_\ell,T^m_k]=2i\sin \alpha_2 (n,\ell;m,k)T^{n+m}_{\ell+k}.\eeq

Now, to construct a symmetry describing two-dimensional system (exotic particles system) in high-dimensional space we suggest the discretization of a continous high-dimensional space, and to move from one 2-dimensional space to another one inside $d$-dimensinal lattice the particles will jump by using the links (Fig.2). Pure mathematically, this behaviour will be supported by noncommuting link operators.\\

First, we define a $d$-dimensional lattice $\Om_d $ as a tensor product of $d$ times two-dimensional lattice $\Om_2$
$$\Om_d=\Om_2 \otimes ...\otimes \Om_2 .$$
This lattice is seen as a discretization of a continous $d$-dimensional space. The translations of exotic particles in this lattice would be realized mathematically through the composition of different elementary translation operators $L_{\ell}$. We define the translation operator on $d$-dimensional lattice as
\beq
T_{p_{i_k}}^{n_{i_\ell }}=R_{i_1 ,...,i_d }^{\frac{n_{i_1 } p_{i_2 }...n_{i_d } }{d}N}L_{\pm i_1}^{n_{i_1 }}L_{\pm i_2}^{p_{i_2 }}...L_{\pm i_d }^{n_{i_d }},
\eeq 
which generate the translation of a massive state; i.e. at its rest (no interaction). We notice that $\ell=1,2,3,...,d$ and $k=2,4,6,...$ if we consider $d$ odd number. $N$ is a number operator, with $i_1 ,...,i_d =1,...,d$. The indices $n_{i_\ell }$ and $p_{i_k}$ indicate the number of times the operators $L_{\pm i_\ell }$ and $L_{\pm i_k }$ respectively were used for the general translation and this fact will fix the reached site, and $R_{i_1 ,...,i_d }=e^{i\epsilon_{i_1 ...i_d}\al}$, with $\epsilon_{i_1 ...i_d}=\frac{1}{2}\sum\limits_{\ell ,k=1}^{d}\epsilon_{i_\ell i_k }$; $\epsilon_{i_\ell i_k }=\pm 1$, as before the sign $\pm$ depends on the rotation direction on $\Om _2 $ the subspace of $\Om _d$.\\

Next, we use the definition (12) of $T_x$ and the supposition (10) of noncommuting link operators $L_\ell $ to find out what commutation relations the operators (15) satisfy. Thus a straightforward computation yields 
\beq
\lbrack T_{p_{i_k}} ^{n_{i_\ell }}, T_{q_{i_k}}^{m_{i_\ell }}\rbrack =\Re(n_{i_\ell },p_{i_k};m_{i_\ell },q_{i_k})T^{n_{i_\ell }+m_{i_\ell }}_{p_{i_k}+q_{i_k}},
\eeq
with the factor $\Re(n_{i_\ell },p_{i_k};m_{i_\ell },q_{i_k})$ is given by the expression
\beq
\bra{ll}
\Re(n_{i_\ell },p_{i_k};m_{i_\ell },q_{i_k})&=R_{i_1 ,...,i_d }^{\Big( \frac{1}{d}\sum\limits_{\al=1}^{d}\Big [ \prod\limits_{
\tiny
\bra{ll}
\be=1\cr
\be\ne \al
\era\rm
}^{d} (n_{i_\be}+p_{i_\be})(m_{i_\al}+q_{i_\al})+\prod\limits_{
\tiny
\bra{ll}
\ga=1\cr
\ga\ne \al
\era\rm
}^{d}(n_{i_\al }+p_{i_\al})(m_{i_\ga}+q_{i_\ga})\Big ] N\Big) }\\ \\
&\times \left\{
R_{i_1 ,...,i_d }^{\sum\limits_{t=1}^{d-1}\frac{m_{i_t}+q_{i_t}}{d-t+1}\prod\limits_{\epsilon=t-1}^{d}(n_{i_\epsilon}+p_{i_\epsilon})N^2 }-R_{i_1 ,...,i_d }^{\sum\limits_{t=1}^{d-1}\frac{n_{i_t}+p_{i_t}}{d-t+1}\prod\limits_{\si=t-1}^{d}(m_{i_\si}+q_{i_\si})N^2 }\right\}
\era
\eeq
where $m_{i_\be},n_{i_\be}=0$ for $\be=2,4,...$ and $q_{i_\be},p_{i_\be}=0$ for $\be=1,3,...,d$. This is applied for all other indices  $t,\si,\epsilon...$.\\

The commutation relations (16) define the $w_{\infty}$-algebra closed by the generators $T_{p_{i_k}}^{n_{i_\ell }}$. This is constructed in a pure mathematical context in high-dimensional lattice which is a composition of 2-dimensional one where exotic particles live. We notice that if $d$ reduces to 2, the factor (17) reduces to $\Re(n_{i_1 },p_{i_2};m_{i_1 },q_{i_2})=2i \phantom{~}sin \al_2 (n,p; m,q)$ and the algebra (16) becomes FFZ algebra as discussed in the begining of this section. Then the fractional statistics particles are kept to move only in two-dimensional space. In the case of $d>2$ the moving is in the whole of high-dimensional space by jumping from one site to another using the links between two-dimentional lattices. We also note that the indices $n$ and $p$ in the equations (15,16,17) should depend on the spin characterizing exotic particles and consequentelly they should be (linearly or not, we don't know for the moment) proportional to the statistical parameter $\al$.

\begin{center}
\begin{picture}(0,0)
\put(-30,0){\line(0,-1){80}}
\put(20,0){\line(0,-1){80}}
\put(-50,-20){\line(1,0){80}}
\put(-50,-70){\line(1,0){80}}

\put(-33,-23){$\bullet$}

\put(20,-20){\vector(-1,0){20}}
\put(-30,-70){\vector(1,0){20}}
\put(20,-70){\vector(0,1){20}}
\put(-30,-20){\vector(0,-1){20}}


\put(-70,-50){\line(0,-1){80}}
\put(-20,-50){\line(0,-1){80}}
\put(-90,-60){\line(1,0){80}}
\put(-90,-110){\line(1,0){80}}

\put(-33,-23){\line(-1,-1){40}}
\put(-33,-73){\line(-1,-1){40}}
\put(17,-23){\line(-1,-1){40}}
\put(17,-73){\line(-1,-1){40}}

\put(-43,-13){$A$}
\put(-58,-39){$L$}

\put(40,-20){$(a)$}
\put(-120,-90){$(b)$}

\put(-33,-23){\vector(-1,-1){10}}

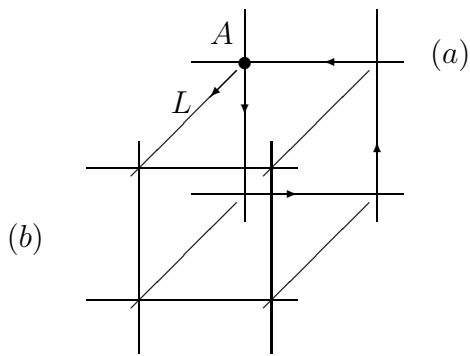
\put(-240,-180){{\bf Fig.2} -- Example from high-dimensional lattice. The exotic particle $A$ is translated}
\put(-200,-200){ from the two-dimensional lattice $(a)$ to $(b)$ using the link $L$ connecting the }
\put(-200,-220){ neighborhood sites.}
\end{picture}

\end{center}

\vspace{7cm}
\section{Conclusion}
\hspace*{.3in}Let us close this note with a short summarize and an interesting aspect of the problem that can be studied in near future. First, we see that the two-dimensional particles are all the time treated in two dimensional space as their own place to live. In this paper we suggested that these particles could live in high-dimensional space by taking into account the way they should translate. We considered the space as a discretized one and we assumed that the particles could be transported from one site to another through the links of the lattice and the braid nature of their motions is saved by the noncommutativity of link operators $L_\ell $. Thus, these latter were the basic objects for the whole work. Second, We remark that the exotic particles should change the orientation of the magnetic flux they carry at the time when they leave one two-dimensional lattice to another inside the whole space; this phenomenon will be generated by a specified link operators. In this paper, we didn't deal with this subject since we didn't need it. We also note that The mathematical tools used to construct the translation algebra describing two-dimensional system in two- and high-dimensional spaces could be very interesting to investigate the quantum computer in which "anyons" play a crucial role \cite{computer}.\\

\end{document}